\documentclass[twocolumn,aps,pra,superscriptaddress,showpacs]{revtex4-1}
\usepackage{amsmath}
\usepackage{graphicx,graphics}
\usepackage{hyperref}
\usepackage{amssymb}
\usepackage{color}
\usepackage{appendix}
\begin{document}
\title{Fermi-Polaron in a driven-dissipative background medium }
\author{Ye Cao }
\thanks{These authors contributed equally to this work.}
\affiliation{School of Physics, Beijing Institute of Technology, Beijing 100081,
China}
\author{Jing Zhou}
\thanks{These authors contributed equally to this work.}
\thanks{corresponding author:zhjing@mail.ustc.edu.cn}
\affiliation{Department of Science, Chongqing University of Posts and Telecommunications,
Chongqing 40006, China}
\date{$\today$}
\begin{abstract}
The study of polaron of an open quantum system plays an important
role in both verifying the effectiveness of approximate many-body
theory and predicting novel quantum phenomenone in open quantum systems.
In a pioneering work, Piazza et al have proposed a Fermi-polaron scheme
with a lossy impurity \cite{Piazza}, which exhibits a novel long-lived
attractive polaron branch in the quantum Zeno limit. However, we would
also run into a counterpart problem that an impurity scatters with
an open quantum bath exciting polarons, which is what we focus in
on. In this work, we conclude the molecular state under the two limits
of vanishing small and infinite large dissipation intensity as well
as the reason why the dissipation range leads to the decrease of the
gap between the molecular state and molecule-hole continuum in the
former case by means of analytically research. The spectrum functions
of molecular and polaron states with different dissipation range and
loss rate are investigated. We find the spectral signals of molecular
and polaron states will both diffuse firstly and then revives as the
dissipation is on the raise. Moreover, it is shown that the attractive
and repulsive polarons show different response to the increasing dissipation
range in our model. At last, we exhibit the polaron energy, residue,
effective mass and two-body decay for mass balanced and imbalanced
systems. Our results might be useful for future cold atom experiment
on open quantum systems.
\end{abstract}
\maketitle

\section{Introduction}

Since the concept of polaron was proposed as quasiparticles arising
from the electron-phonon interaction in the 1930s \cite{Landau},
the mechanism of a mobile impurity interacting with its enviroment
has gained widespread concern. It plays a key role in understanding
the low-energy behavior of complex systems, e.g., cuprate superconductors
\cite{Lee}, colossal magnetoresistive manganites \cite{Mannella},
$^{3}\text{He}-^{4}\text{He}$ mixtures \cite{Baym}, and promising
organic semiconductors \cite{Gershenson}, to name a few.

Due to the excellent flexibility, the ultracold atomic system has
attracted extensive attention in recent years \cite{Bloch,Chin}.
Especially, the progress of the study on polaron continues to spring
up in our times, both experimentally \cite{Zwierlein1,Thomas,Grimm,Kohl,Demler1,Jin,Christensen,Roati,Zwierlein2,Zwierlein3,Sagi}
and theoretically \cite{chevy1,stringari,chevy2,prokofev1,stoof,Giraud,Zwerger,Bruun,Parish1,Demler2,Parish2,Houcke,Schmidt,Kinnunen,Sarma,Pollet,Parish3,Hui1,prokofev2,Hui2,Uchino,Hui3,Parish4,Hui4,Parish5,Massignan,Zhai1}.
However, these investigations are restricted to isolated systems.
With the development of quantum physics, the open quantum systems
have become more and more significant. The nonnegligible coupling
of some specific systems with their environments, the diseconomy and
difficulty of calculating a large isolated system and removing overwhelming
useless information to obtain the specific degree of freedom which
we pay attention to, are a couple of reasons why the theory of open
quantum system plays such a crucial role.

Further more, dissipation has been proved to be pivotal in the production
of novel quantum phenomenas. Many breakthroughs have been made in
the theoretical and experimental areas. For instance, dissipation
can induce strong correlation in ultracold Bosonic atoms \cite{Durr}
and controll phase transition from the Mott insulator to the superfluid
\cite{Takahashi}, slow down the relaxation of many-body states to
an algebraic law \cite{Gerbier}, engineer tunable local loss in
a synthetic lattice of momentum state \cite{Gadway}, and bring about
$\mathcal{PT}$ -symmetry-breaking transitions that makes a $\mathcal{PT}$
-symmetric broken phase which is beyond the quantum Zeno effect comes
into being \cite{Luo}. From another perspective, as an effective
theory to describe the system undergoing dissipation, the non-Hermitian
Hamiltonian approach has aroused great interest in recent years \cite{Ashida1},
inspiring the research from the aspects of skin effect, dynamics,
band theory, topological phase-transition, non-Hermitian linear response
theory, non-Hermitian semimetal and dissipation-facilitated molecule
and so on.

Nevertheless, to the best of our knowledge, the works about dissipative
polaron \cite{Piazza,Zhou,Sun}are still rare. On the one hand, benefiting
from the simplicity, polaron physics is suitable for checking the
consistency of many-body theory and experimental results. As a pioneering
work \cite{Piazza}, Piazza et. al. have considered a polaron scheme
that the impurity atom bears driving and dissipation which do not
come from the bath but from other environment degrees, where they
found quantum Zeno effect in the impurity spectrum function. On the
other hand, another important mission of a polaron is to act as a
probe of its bath, e.g., a complex open quantum system which we make
concern of. Based on the importance of many-body theory for a quantum
open system and the thirst for knowledge of open quantum system which
is difficult to calculate, it is worthwhile to carry out new polaron
studies, in the scenario of which the bath is dissipated and driven
by its own environment.

In this work, we study a new polaron system whereby the bath turns
to be Fermi gases suffering external driving and dissipation. In order
to take the effect of quantum jump term of the Lindblad master equation
(LMEQ) into account, we commence the study with the total action function
of the impurity and background gases. Employing the non-equilibrium
Green's function (GF) method, we present the spectrum responses of
molecular and polaron states under different system configurations.
We also calculate quasiparticle parameters in both mass balanced and
imbalanced polaron setup, which can be realized in the experiments.
Two analytic results are obtained. Firstly, for the vanishingly small
dissipation, we explain the reason why the gap between molecular state
and molecule-hole continuum decreases with the increase of dissipation
range. Secondly, for the infinite dissipation, we obtain the dispersion
of bound state as well as the threshold of molecule-hole continuum.
We find the resonant peaks of molecular and polaron states become
diffuse only under moderate dissipation. On the contrary, they become
well-defined with infinite dissipation. In other words, under this
limit, the dissipation range only plays a role in reconstructing the
Fermi surface. This mechanism is consistent with the Zeno effect described
in \cite{Ueda,Piazza}. The residue, effective mass and two-body
decay show nonmonotonic characters due to the interplay between the
intrinsic energy scale of the system and the measurement frequency
from the environment. However, they all tend to respective saturation
values under the Zeno limit.

The paper is organized as follows: in Sec. \ref{sec:Model-and-LMEQ},
we introduce our model and Keldysh diagrammatic approach to our dissipative
polaron problem. In Sec. \ref{sec:Numerical-Results}, we discuss
the numerical results of the molecular and the polaron states. In
Sec. \ref{sec:Conclusion-and-outlook}, we make a conclusion and give
some remarks and outlook on this research topic.

\section{\label{sec:Model-and-LMEQ}Model and Analytical Approach}

\subsection{Model Hamiltonian and LMEQ}

Let us consider a system composed of a fermionic impurity and a two-dimensional
open Fermi bath interacting with it. The Hamiltonian is expressed
as
\begin{equation}
\mathcal{H}=\mathcal{H}_{\text{imp}}+\mathcal{H}_{\text{bath}}+\mathcal{H}_{\text{int}},\label{eq:total-Hamiltonian}
\end{equation}
where $\mathcal{H}_{\text{imp}}=\sum_{\mathbf{k}}\varepsilon_{c}(\mathbf{k})c_{\mathbf{k}}^{\dagger}c_{\mathbf{k}}$
and $\mathcal{H}_{\text{bath }}=\sum_{\mathbf{k}}\varepsilon_{f}(\mathbf{k})f_{\mathbf{k}}^{\dagger}f_{\mathbf{k}}$
are the kinetic Hamiltonians of impurity and bath, the dispersions
of which are $\varepsilon_{c}(\mathbf{k})=\mathbf{k}^{2}/2m_{c}$
and $\varepsilon_{f}(\mathbf{k})=\mathbf{k}^{2}/2m_{f}$, respectively.
$\mathcal{H}_{\text{int}}=g\int d\mathbf{r}c^{\dagger}(\mathbf{r})c(\mathbf{r})f^{\dagger}(\mathbf{r})f(\mathbf{r})$
is the $s$-wave contact interaction between the impurity and the
bath. The dynamics of the bath should be describe by a LMEQ,
\begin{equation}
\partial_{t}\rho_{\text{bath}}=-i[\mathcal{H}_{\text{bath}},\rho_{\text{bath}}(t)]+\mathcal{L}_{d}\rho_{\mathrm{bath}}(t),\label{eq:general-LMEQ}
\end{equation}
where the $\mathcal{L}_{d}$ is the dissipative Lindblad operator,
whose effect on the density operator is $\mathcal{L}_{d}\rho_{\text{bath}}=\sum_{\mathbf{k}}\{\gamma(\mathbf{k})D[f_{\mathbf{k}}]+\Omega(\mathbf{k})P[f_{\mathbf{k}}]\}\rho_{\text{bath}}$,
describing loss and reinjection of particles with rate of $\gamma(\mathbf{k})$
and $\Omega(\mathbf{k})$. The two incoherent processes are induced
by the superoperators, $D[f_{\mathbf{k}}]\rho_{\text{\text{bath}}}\equiv f_{\mathbf{k}}\rho_{\text{bath}}f_{\mathbf{k}}^{\dag}-\frac{1}{2}\{f_{\mathbf{k}}^{\dag}f_{\mathbf{k}},\rho_{\text{bath}}\}$
and $P[f_{\mathbf{k}}]\equiv-D[f_{\mathbf{k}}]+D[f_{\mathbf{k}}^{\dag}]$,
where we have chosen a specific pumping behavior for simplicity.
The action of $D[f_{\mathbf{k}}]$ operating on the density matrix
$\rho_{\text{bath}}$ results in two processes, one is the continuous
nonunitary evolution coming from the anticommutator term, the other
one is quantum jump produced by $f_{\mathbf{k}}\rho_{\text{bath}}f_{\mathbf{k}}^{\dag}$.
The former describes the continuous losses of energy and information
in the process of decoherence with environment, and the later represents
the continuous measurement of the system.

For our model, there is not interaction between the bath atoms, so
the matrix representing of the density operator is block diagonal
in bases of the Fock states for each momentum. Further more, the bath
is supposed to keep in a steady state where the loss and pump are
already in equilibrium, therefore the matrix elements of the density
matrix is time independent that can be find out through the LMEQ as
$\rho_{\mathbf{k}}^{00}=1-\Omega(\mathbf{k})/\gamma(\mathbf{k})$,
$\rho_{\mathbf{k}}^{01}=0$, $\rho_{\mathbf{k}}^{11}=\Omega(\mathbf{k})/\gamma(\mathbf{k})$.
In addition, the underlying probability annotation of the density
operator requires that the ratio $\Omega\left(\mathbf{k}\right)/\gamma(\mathbf{k})$
must be confined in the range of $[0,1]$. For each momentum, the
average particle number is determined by this ratio, the momentum
distribution of which further establishes the average total particle
number.

\subsection{$T$-matrix method with Non-equilibrium Green's function}

There are two accepted ways to deal with the polaron system, variational
ansatz method and many-body $T$-matrix method, which have been proved
to be equivalent by Chevy in 2007 \cite{Chevy}. The former is intuitive
and has been extended to the dynamic version \cite{Parish6,Parish7}.
Meanwhile, the later is a little more complicated. We need to calculate
the self energy perturbatively and then obtain the spectrum function
as well as quasiparticle parameters. However, the $T$-matrix approach
can be conveniently constructed by the non-equilibrium GF to give
a complete description of the open quantum system taking into account
all the effects of the LMEQ without ignoring the quantum jump term
\cite{kamenev}, which is the method we adopt in this work.

We start from the full partition function of the system,
\begin{equation}
\mathcal{Z}=\int_{C}D[\bar{c},c,\bar{f},f]e^{iS_{\text{imp}}+iS_{\text{bath}}+iS_{\text{int}}},
\end{equation}
where $C$ is the Keldysh integral contour. Performing the Fermi Keldysh
rotation, the bare impurity action reads
\begin{eqnarray}
S_{\text{imp}} & = & \sum_{k}\overline{c}_{\mu}(k)G_{0,c}^{-1,\text{\ensuremath{\mu\nu}}}(k)c_{\nu}(k),
\end{eqnarray}
where the $\mu,\nu=1,2$ are Keldysh indices and the Einstein summation
convention of repeated indices is connoted here and thereinafter.
The matrix of $G_{0,c}^{-1}$ is upper triangular, so we can straightforwardly
read out the bare retarded and advanced GFs
\begin{equation}
G_{0,c}^{R/A}(\mathbf{k},\omega)=[\omega-\varepsilon_{c}(\mathbf{k})\pm i0^{+}]^{-1}.
\end{equation}
According to the fluctuation-dissipation relation (FDR) for steady
state, the bare Keldysh GF can always be expressed with the corresponding
retarded and advanced GFs ,
\begin{equation}
G_{0,c}^{K}(\mathbf{k},\omega)=F_{c}^{eq}(\omega)[G_{0,c}^{R}(\mathbf{k},\omega)-G_{0,c}^{A}(\mathbf{k},\omega)].
\end{equation}
In the impurity limit, where the density of impurity atoms is vanishing
small, we have $F_{c}^{eq}=1$.

The GFs for the bath are some complicated. The retarded, advanced
and Keldysh GFs should be obtained simultaneously by reverting the
matrix of $G_{0,f}^{-1}$, which can be extracted from the dissipative
bath action with a Keldysh rotation applied (see Appendix).
\begin{equation}
G_{0,f}^{R/A}(\mathbf{k},\omega)=[\omega-\varepsilon_{c}(\mathbf{k})\pm\frac{i\gamma(\mathbf{k})}{2}]^{-1},
\end{equation}
\begin{equation}
G_{0,f}^{K}(\mathbf{k},\omega)=F_{0,f}(\mathbf{k},\omega)[G_{0,f}^{R}(\mathbf{k},\omega)-G_{0,f}^{A}(\mathbf{k},\omega)],
\end{equation}
where the distribution function $F_{0,f}(\mathbf{k},\omega)=1-2\eta(\mathbf{k})$
and $\eta(\mathbf{k})=\Omega(\mathbf{k})/\gamma(\mathbf{k})$ is the
average particle number for momentum $\mathbf{k}$.

Next, we introduce an auxiliary molecular field $\Delta$ to decouple
the four-operator interaction term in the total action. By performing
a Hubbard-Stratonovich transformation, the interaction part of the
action becomes
\begin{equation}
S_{\Delta,c,f}=\int_{C}dx(-\overline{\Delta}fc-\overline{c}\overline{f}\Delta),
\end{equation}
and further changes into
\begin{equation}
S_{\Delta,c,f}=-\int dx\frac{1}{\sqrt{2}}(f_{\mu}\overline{\Delta}_{a}\gamma_{\mu\nu}^{a}\sigma_{\nu\eta}^{x}c_{\eta}+\text{h.c.}),
\end{equation}
with a successive Keldysh rotation to achieve a basis transformation,
where $\mu,\nu,\eta=1,2$, and $a=q,cl$ are fermion and boson Keldysh
indices respectively. The elements of $\gamma^{a}$ are $\gamma^{cl}=I$
and $\gamma^{q}=\sigma^{x}$, where $I$ is the identity matrix and
$\sigma^{x}$ is the $x$ component of Pauli matrices. To calculate
the self-energy, we integrate out the fermionic bath degrees of freedom
and obtain the induced interaction action for molecule and impurity,
\begin{eqnarray}
S_{\Delta,c} & = & -\frac{1}{2}\int dxdx'\overline{\Delta}_{a}(x)\gamma_{\mu\nu}^{a}\sigma_{\nu\eta}^{x}c_{\eta}(x)G_{0,f}^{\mu\tau}(x,x')\nonumber \\
 &  & \times\overline{c}_{\alpha}(x')\sigma_{\alpha\beta}^{x}\gamma_{\beta\tau}^{b}\Delta_{b}(x').
\end{eqnarray}
Under the framework of non-self-consistent $T$-matrix method \citep{Piazza},
accurate to first order, the real space self-energy of the molecule
and impurity are solved out as,

\begin{equation}
\Sigma_{\Delta}^{ab}(x,x')=\frac{i}{2}\gamma_{\mu\nu}^{a}\sigma_{\nu\eta}^{x}G_{0,f}^{\mu\tau}(x,x')G_{0,c}^{\eta\alpha}(x,x')\sigma_{\alpha\beta}^{x}\gamma_{\beta\tau}^{b},\label{eq:self-energy-molecule}
\end{equation}
for molecule, and
\begin{equation}
\Sigma_{c}^{\alpha\eta}(x,x')=\frac{i}{2}\sigma_{\alpha\beta}^{x}\gamma_{\beta\tau}^{b}G_{\Delta}^{ba}(x',x)G_{0,f}^{\mu\tau}(x',x)\gamma_{\mu\nu}^{a}\sigma_{\nu\eta}^{x}.\label{eq:self-energy-impurity}
\end{equation}
the self-energy of molecule and impurity can be expressed by the corressponding
retarded, advanced and Keldysh GFs respectively.

\subsection{The retarded self-energy and spectrum function}

To obtain the spectrum functions and quasiparticle parameters, we
need to get the GFs of molecule and impurity, which is essentially
to obtain the corresponding retarded self-energy. In the steady state,
the system is time- and space-translational invariant, so all the
real-space GFs and self-energy can be written out by their Fourier
transforms for the relative coordinates, which is more convenient
to discuss the quasiparticle properties. Setting $a=q$ and $b=cl$
in Eq. \eqref{eq:self-energy-molecule} and taking the Fourier transform,
we obtain the momentun-space retarded self-energy of the molecule
state,
\begin{multline}
\Sigma_{\Delta}^{R}(q)=\frac{i}{2V}\sum_{p_{1}}\Big\{ G_{0,f}^{R}(q-p_{1})G_{0,c}^{K}(p_{1})\\
+G_{0,f}^{K}(q-p_{1})G_{0,c}^{R}(p_{1})\Big\}.
\end{multline}
Performing the contour integral for frequency, the retarded self-energy
of molecule becomes
\begin{align}
\Sigma_{\Delta}^{R}(q) & =\frac{1}{V}\sum_{\mathbf{p}_{1}}\frac{1-\eta(\mathbf{p}_{1})}{\omega-\varepsilon_{f}(\mathbf{p}_{1})-\varepsilon_{c}(\mathbf{q}-\mathbf{p}_{1})+i\frac{\gamma(\mathbf{p}_{1})}{2}},\label{eq:molecule-retarded-self-energy}
\end{align}
and the retarded GF of molecule is written as
\begin{equation}
G_{\Delta}^{R}(\mathbf{q},\omega)=\frac{1}{\frac{1}{g}-\Sigma_{\Delta}^{R}(\mathbf{q},\omega)+i0^{+}}.
\end{equation}
Following the similar procedure, the retarded self-energy of impurity
is expressed as
\begin{eqnarray}
\Sigma_{c}^{R}(k) & = & -\frac{i}{2V}\sum_{p_{1}}G_{\Delta}^{K}(k_{+})G_{0f}^{A}(p_{1})+G_{\Delta}^{R}(k_{+})G_{0f}^{K}(p_{1})\nonumber \\
 & = & \frac{1}{V}\sum_{\mathbf{p}_{1}}\eta(\mathbf{p}_{1})G_{\Delta}^{R}(\mathbf{p}_{1}+\mathbf{k},\varepsilon_{f}(\mathbf{p}_{1})+\omega),
\end{eqnarray}
where $k_{+}=p_{1}+k$. Then the retarded GF of the impurity is given
by
\begin{equation}
G_{c}^{R}(k)=\frac{1}{\omega-\varepsilon_{c}(\mathbf{k})-\Sigma_{c}^{R}(k)+i0^{+}}.\label{eq:retarded-GF-of-impurity}
\end{equation}

The spectrum functions of molecule and impurity are written as
\begin{eqnarray}
A_{\Delta}(q) & = & -\frac{1}{\pi}\text{Im}[G_{\Delta}^{R}(q)],\\
A_{c}(k) & = & -\frac{1}{\pi}\text{Im}[G_{c}^{R}(k)].
\end{eqnarray}
The pole of the retarded GF of impurity determines the polaron energy,
that is
\begin{equation}
E_{p}(\mathbf{k})=\varepsilon_{c}(\mathbf{k})+\text{Re}\Sigma_{c}^{R}(\mathbf{k},E_{p}).\label{eq:polaron-energy}
\end{equation}
For our system, $E_{p}$ always increases as $|\mathbf{k}|$ on the
raise, so only the solutions at zero momentum are considered and the
equation above develops into $E_{p}=\text{Re}\Sigma_{c}^{R}(0,E_{p})$.
According to Landau Fermi liquid theory, a polaron state behaves like
a quasiparticle with collective energy shift $E_{p}$, residue $Z$,
effective mass $m_{c}^{*}$ , and the two-body decay rate $\gamma_{+}$,
so the asymptotic behavior of the GF in the zero momentum limit takes
the form
\begin{equation}
G_{c}^{R}(\mathbf{k},\omega)=\frac{Z}{\omega-\frac{\hbar^{2}\mathbf{k}^{2}}{2m_{c}^{*}}-E_{p}+i\gamma_{+}}.\label{eq:Fermi-linquid-GF}
\end{equation}
Expanding the retarded self-energy in Eq. \eqref{eq:retarded-GF-of-impurity}
to first order at the pole and Comparing the GF with that in Eq. \eqref{eq:Fermi-linquid-GF},
we have
\begin{eqnarray}
Z & = & \frac{1}{1-\frac{\partial\mathrm{Re}\Sigma_{c}^{R}(0,\omega)}{\partial\omega}|_{\omega=E_{p}}},\label{eq:Z}\\
\frac{m_{c}^{*}}{m_{c}} & = & \frac{1-\frac{\partial\mathrm{Re}\Sigma_{c}^{R}(0,\omega)}{\partial\omega}|_{\omega=E_{p}}}{1+\frac{\partial\mathrm{Re}\Sigma_{c}^{R}(\varepsilon_{c}(\mathbf{k}),E_{p})}{\partial\varepsilon_{c}(\mathbf{k})}|_{k=0}},\label{eq:m}\\
\gamma_{+} & = & -2Z\text{Im}\Sigma_{c}^{R}(0,E_{p}).\label{eq:gamma}
\end{eqnarray}
For two-dimensional Fermi gases, the above bare interaction strength
$g$ should be renormalized by the two-body binding energy $E_{B}$
as
\begin{equation}
\frac{1}{g}=-\sum_{k}\frac{1}{E_{B}+\varepsilon_{r}},
\end{equation}
where $\varepsilon_{r}=\frac{\hbar^{2}k^{2}}{2m_{r}}$, and $m_{r}=\frac{m_{c}m_{f}}{m_{c}+m_{f}}$
is the reduced mass.

\section{\label{sec:Numerical-Results}numerical results and discussion}

\subsection{Dressed molecule state}

The molecular state plays an vital role in the scenario of the interaction
between the impurity and the bath, not only because the introduction
of auxiliary field can easily decouple the interaction potential,
but also because its retarded GF is just the $T$-matrix characterizing
the moving impurity scattered by the bath fermions. From the formula
derived above, we could conclude that all the GFs depend on the $\gamma(\mathbf{k})$
and the ratio of $\text{\ensuremath{\eta(\mathbf{k})=}}\Omega(\mathbf{k})/\gamma(\mathbf{k})$.
For simplicity, we limit the driving and dissipation to the same range
depicted by a cutoff momentum $k_{r}$, within which both the $\gamma(\mathbf{k})=\gamma_{0}$
and $\eta(\mathbf{k})=\eta$ are nonzero constants and zero elsewhere.
In addition, unless otherwise specified, we set $m_{c}=m_{f}=1/2$
and $k_{F}=1$ in the calculations below.

\begin{figure*}
\includegraphics[width=1\linewidth]{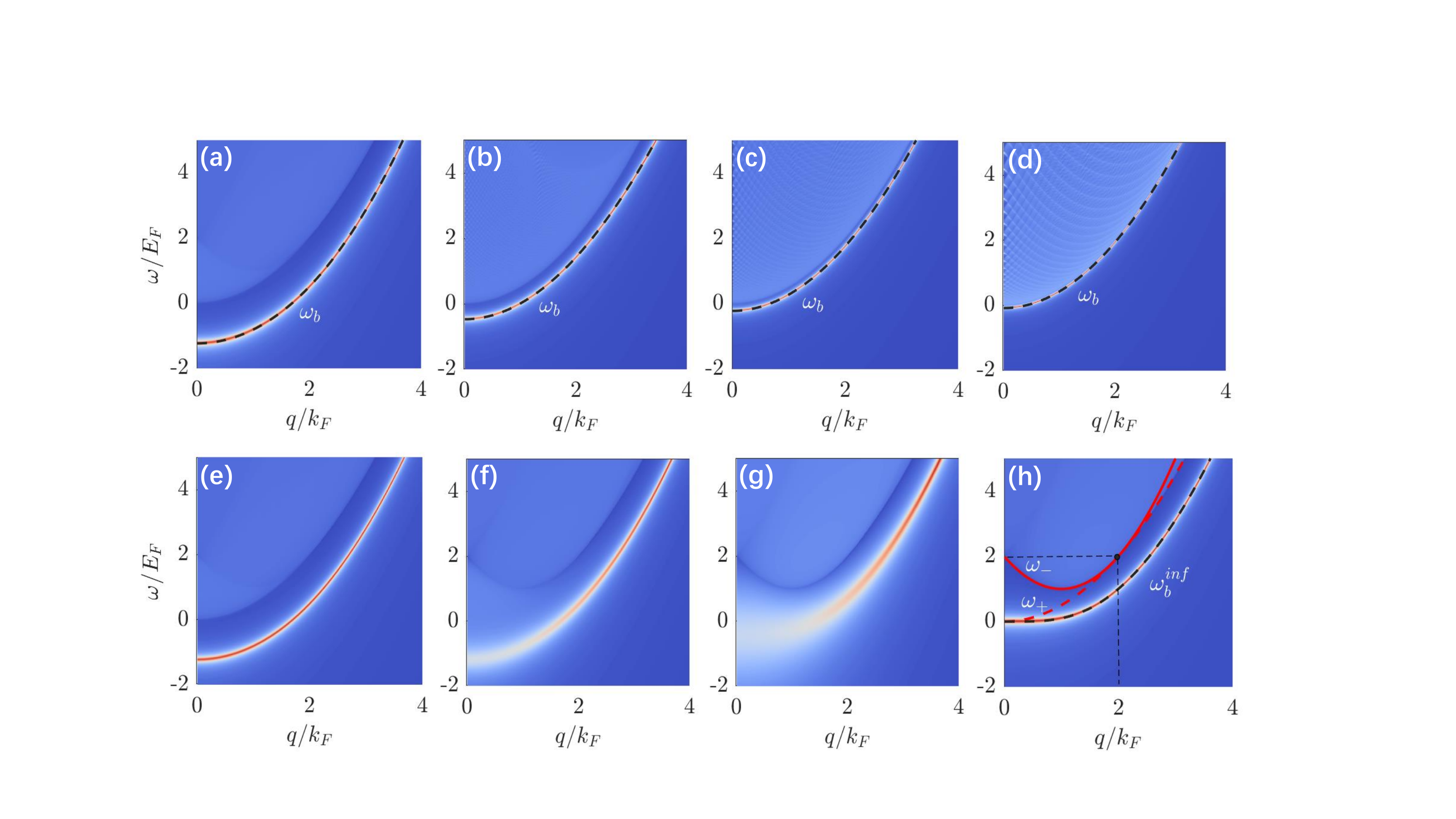}\hfill{}
\caption{\label{Fig: molecule-dispersion}(Color online) Molecular spectral
function $A_{\Delta}$ with different dissipation cutoff momentum
$k_{r}$ {[}(a), $k_{r}/k_{F}=1.0$; (b), $k_{r}/k_{F}=2.0$; (c),
$k_{r}/k_{F}=3.0$; (d), $k_{r}/k_{F}=5.0${]} for a small loss rate
$\gamma_{0}/E_{F}=0.01$, and with different dissipation strength
$\gamma_{0}$ {[}(e), $\gamma_{0}/E_{F}=0.01$; (f), $\gamma_{0}/E_{F}=1.0$;
(g), $\gamma_{0}/E_{F}=5.0$; (h), $\gamma_{0}/E_{F}=30${]} for $k_{r}/k_{F}=1.0$.
Other parameters are $\eta=0.5$, $E_{B}/E_{F}=2$. In addition, the
black dashed lines in (a)-(d) are the analytical results of molecular
bond state $\omega_{b}$ with $\gamma_{0}=0^{+}$ and the same other
parameters in their respective subplots, which are in perfect agreement
with the numerical results at small $\gamma_{0}$. In (h), we show
the analytical solution of the lower bound of molecule-hole continuum
at infinit dissipation marked by $\omega_{-}$ (red solid line) and
$\omega_{+}$ (red dashed line) to the left and right of their intersection
(black dot) respectively (A detailed explanation is provided later
in this section). Meanwhile, the $\omega_{b}^{inf}$ (black dashed
line) illustrates the analytical solution of molecular bound state
under the same conditions. These analytical results are also in line
with the numercial results at large $\gamma_{0}$. }
\end{figure*}

We start by analysing the parameter regime where the dissipation is
vanishing small. On this condition, the self-energy of the molecular
field can be separated into two parts as $\Sigma_{\Delta}^{R}=\Sigma_{\Delta}^{R,1}+\Sigma_{\Delta}^{R,2}$,
where the first part is $\Sigma_{\Delta}^{R,1}=1/V\sum_{\mathbf{p}_{1}}1/[(\omega-\varepsilon_{f}(\mathbf{p_{1}})-\varepsilon_{c}(\mathbf{q}-\mathbf{p_{1}})+i0^{+})]$
and the second part is $\Sigma_{\Delta}^{R,2}=-\eta/V\sum_{|\mathbf{p_{1}}|<k_{r}}1/[\omega-\varepsilon_{f}(\mathbf{p_{1}})-\varepsilon_{c}(\mathbf{q}-\mathbf{p_{1}})+i0^{+}]$.
Here, we only focus on the molecular bound state, so the energy must
be less than the kinetic energy of the center of mass of a molecule,
that is $\omega<\min_{\mathbf{p_{1}}}[\varepsilon_{f}(\mathbf{q}/2+\mathbf{p}_{1})+\varepsilon_{c}(\mathbf{q}/2-\mathbf{p}_{1})]=q^{2}/2M$.
In this case, the two integrands in $\Sigma_{\Delta}^{R,1}$ and $\Sigma_{\Delta}^{R,2}$
do not have imaginary parts. Setting the upper limit of integral of
$\mathbf{p}_{1}$ to $\Lambda$ in $\Sigma_{\Delta}^{R,1}$, we conclude
that

\begin{equation}
\Sigma_{\Delta}^{R,1}=-\frac{1}{8\pi}\ln|\frac{\frac{\Lambda^{2}}{\widetilde{\mu}}-\widetilde{\omega}+u_{q}}{u_{q}-\widetilde{\omega}}|,
\end{equation}
where we have used the notations $\widetilde{\mu}=k_{r}^{2}$, $x=p_{1}^{2}/\widetilde{\mu}$,
$\widetilde{\omega}=\omega/(2\widetilde{\mu})$, $\omega_{q}^{*}=q^{2}/2M$
and $u_{q}=\omega_{q}^{*}/(2\widetilde{\mu})$ to simplify the expression
above. In the meantime, the $\Sigma_{\Delta}^{R,2}$ evolves into
\begin{multline}
\Sigma_{\Delta}^{R,2}=\int_{0}^{1}dx\int_{0}^{2\pi}d\theta\frac{\eta/(16\pi^{2})}{x+2u_{q}-\widetilde{\omega}-2\sqrt{xu_{q}}\cos{\theta}}.
\end{multline}
Using the integral identity
\begin{equation}
\int_{0}^{2\pi}d\theta\frac{1}{a+b\cos{\theta}}=\frac{\pi\text{sign}(a)\Theta(a^{2}-b^{2})}{\sqrt{a^{2}-b^{2}}},
\end{equation}
we can figure out the second part of retarded self-energy of the molecular
field and then obtain the inverse of the retarded GF of the molecular
field
\begin{multline}
G_{\Delta}^{R,-1}(\mathbf{q},\omega)=\frac{1}{g}-\Sigma_{\Delta}^{R}(\mathbf{q},\omega)=\frac{1}{8\pi}\ln|\frac{E_{B}/(2\widetilde{\mu})}{u_{q}-\widetilde{\omega}}|\\
-\frac{\eta}{16\pi}\ln[\frac{\sqrt{\frac{(\widetilde{\omega}-1)^{2}}{4}+u_{q}(u_{q}-\widetilde{\omega})}-\frac{\widetilde{\omega}-1}{2}}{u_{q}-\widetilde{\omega}}].
\end{multline}
Solving the pole of $G_{\Delta}^{R}(\mathbf{q},\omega)$, we can obtain
the molecular bound state shown as the black dashed lines in Fig.
\eqref{Fig: molecule-dispersion} (a)-(d), which are in accord with
the corressponding numerical results obtained under small $\gamma_{0}$
and the same other parameters. It is interesting that the gap between
the boundary of molecule-hole continuum and the molecular bound state
decreases when the dissipation cutoff momentum $k_{r}$ is on the
raise, which can be understood by the composition of retarded self-energy
when the dissipation is infinitesmall. In this instance, $\Sigma_{\Delta}^{R,1}$
is the self-energy of the vacuum scattering (impurity scatters with
a single bath fermion) without dissipation. $\Sigma_{\Delta}^{R,2}$
comes from a finite density of bath gases, which represents the contribution
of dissipative bath fermions to the self-energy by participating in
the formation of molecular state. For the vaccum scattering, the $T$-matrix
is known as
\begin{equation}
T_{0}^{-1}(\mathbf{q},\omega)=-\frac{m_{r}}{2\pi}[ln(\frac{E}{E_{B}})-i\pi],
\end{equation}
where $E=\omega-\mathbf{q}^{2}/2M+i0^{+}$ and $M=m_{c}+m_{f}$. It
is clear that $T_{0}$ has a pole at $\omega^{0}=\mathbf{q}^{2}/2M+i0^{+}-E_{B}$.
When the background participates in the scattering process, the equation
of bound state energy develops into
\begin{equation}
T_{0}^{-1}(\mathbf{q},\omega^{*})=\Sigma_{\Delta}^{R,2}(\mathbf{q},\omega^{*}).\label{eq:bound-state-equation}
\end{equation}
Since the energy of bound state must be less than kinetic energy of
center of mass, then we obtain immediately that $\ln[(\omega^{0}-\mathbf{q}^{2}/2M)/(\omega^{*}-\mathbf{q}^{2}/2M)]>0$,
claiming $\omega^{\ast}>\omega_{0}$. Moreover, the right hand side
of Eq. \eqref{eq:bound-state-equation} grows monotonically with $k_{r}$,
so as the $\omega^{*}$, which explains why the gap between the energy
of bound state and that of boundary of molecule-hole continuum narrows
with $k_{r}$.

In the presence of dissipation, the molecular bound state shows obvious
non-monotonic behavior. Under moderate dissipation strength, the spectral
function presents a certain degree of broadening, as shown in Fig.
\ref{Fig: molecule-dispersion}(f), (g). However, it becomes well
defined again, when $\gamma_{0}$ grows to a large enough value, e.g.,
$\gamma_{0}=30E_{F}$ as shown in Fig. \ref{Fig: molecule-dispersion}(h),
indicating the revival of bound state. It is that the competition
between the increase of argument of the self-energy by the growth
of dissipation and the decrease of its modulus by the same reason
prompts this non monotonicity. For each channel within $k_{r}$, when
the dissipation increases slightly from zero, the imaginary part of
the self-energy increases significantly, which is particularly obvious
for small momentum channels. However, when the dissipation is soaring,
the contribution of the summation terms to the self-energy in all
the dissipative channels will eventually be suppressed by the divergent
denominator in Eq. \eqref{eq:molecule-retarded-self-energy}, resulting
in the attenuation of all the channels within $k_{r}$.

Especially, if $\gamma_{0}$ is infinite, the retarded molecular selfenergy
is given by
\begin{equation}
\Sigma_{\Delta}^{R}(\mathbf{q},\omega)=\sum_{|\mathbf{p_{1}}|>k_{r}}\frac{1/V}{\omega-\varepsilon_{f}(\mathbf{p_{1}})-\varepsilon_{c}(\mathbf{q}-\mathbf{p_{1}})+i0^{+}}
\end{equation}
In this limit, the dissipative boundary takes a similar role as the
Fermi surface. In other words, effective ``Pauli blocking'' is reproduced
in this situation where the $T$-matrix develops into
\begin{equation}
T(q,\omega)=T_{0}(\frac{z}{2}\pm\frac{1}{2}\sqrt{(z-\varepsilon_{f})^{2}-4\varepsilon_{f}(q)\frac{k_{r}^{2}}{2m}}),\label{eq:t-matrix-infinity}
\end{equation}
specifing $z=\omega-\frac{\hbar^{2}k_{r}^{2}}{2m_{r}}+i0^{+}$, and
$\pm=\text{sign}\{\text{Re}[z-\varepsilon_{f}(q)]\}$. Proceed to
the next step, we analytically solve the dispersion of the bound state
out as
\begin{equation}
\omega_{b}^{inf}(\mathbf{q})=\frac{\frac{1}{4}\varepsilon_{f}(\mathbf{q})(\varepsilon_{f}(\mathbf{q})-4\frac{k_{r}^{2}}{2m})-E_{b}^{2}}{E_{B}+\frac{1}{2}\varepsilon_{f}(\mathbf{q})}+\frac{k_{r}^{2}}{2m_{r}},\label{eq:energy-infity}
\end{equation}
which is ploted in Fig. \ref{Fig: molecule-dispersion}(h) with black
dashed line and is very consistent with the numerical result under
large $\gamma_{0}$. Meanwhile, it is the minimum value of $\frac{q^{2}}{2M}+\frac{p_{1}^{2}}{m}$
with the confinement $|\frac{\mathbf{q}}{2}+\mathbf{p_{1}}|>k_{r}$
that determine the threshold of the molecule-hole continuum. This
optimization problem needs to be discussed in different situations.
On the one hand, if $k_{r}<\frac{q}{2}$, then $\omega_{+}(q)=\frac{q^{2}}{2M}$;
on the other hand, if $k_{r}>\frac{q}{2}$, then $\omega_{-}(q)=\frac{(q-k_{r})^{2}+k_{r}^{2}}{2m}$.
The two curves intersect at the $(2k_{r},2k_{r}^{2})$ as shown in
Fig. \ref{Fig: molecule-dispersion}(h). When the dissipation is infinite,
the behavior of molecular state is the same as that in the scattering
of impurity and Fermi sphere. In fact, this phenomenon can also be
interpreted as a Zeno effect, where the dissipation processes can
be viewed as continuous measurement \cite{Piazza}. A similar scenario
has been found by Ueda et.al.\cite{Ueda} in a non-Hermitian BCS
superfluid with complex-valued interaction, which is due to inelastic
scattering between fermions. In their work, the superfluid gap never
collapses but it is enhanced by the dissipation as a result of Zeno
effect. Actually, in the impurity limit, our molecular bound state
can be viewed as a counterpart of the population-balanced Fermi superfluid
state.

\begin{figure}
\includegraphics[width=1\linewidth]{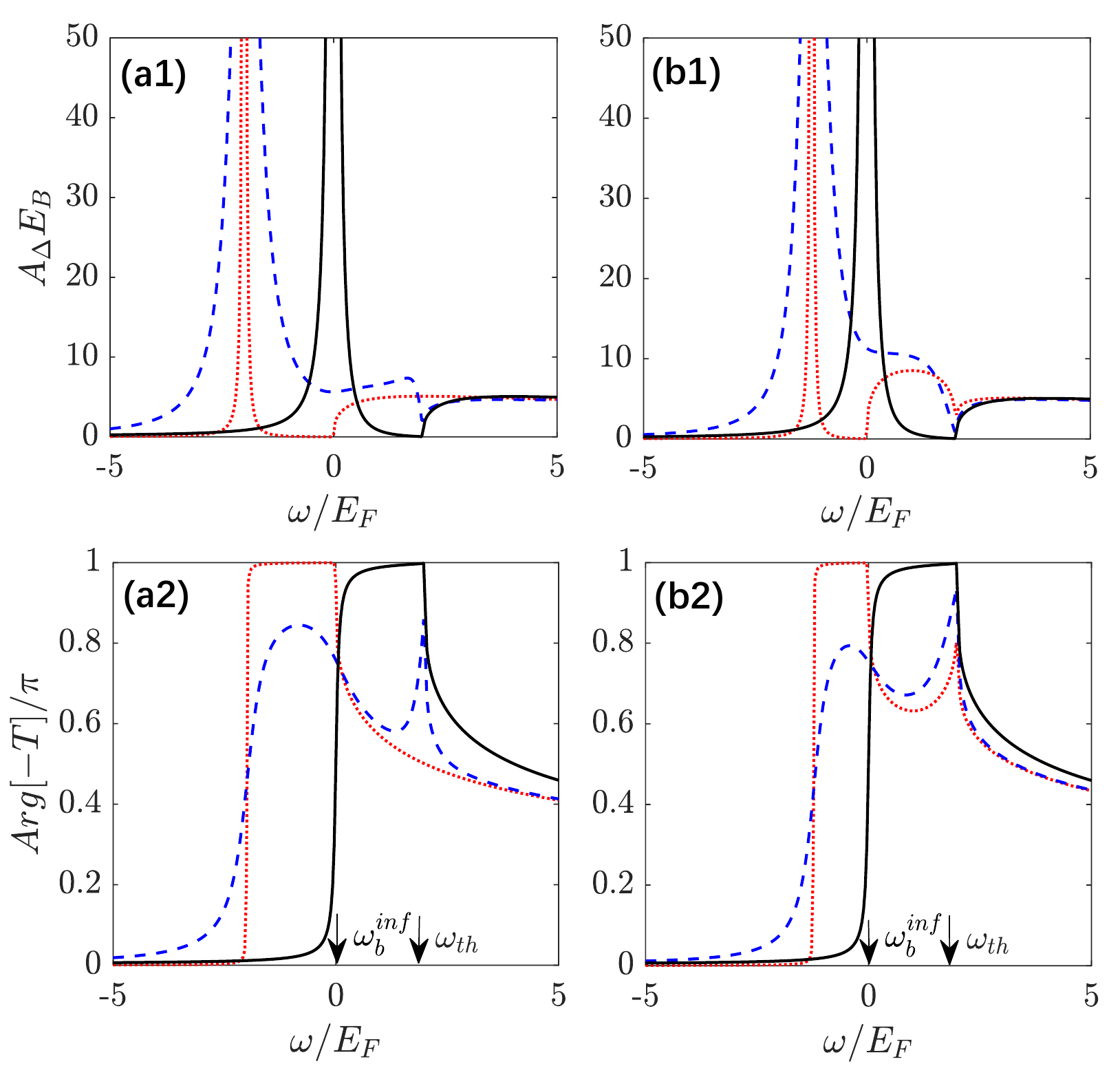}\caption{\label{Fig: molecule-phase}(Color online) Spectrum function of the
molecular bond state at zero momentum, i.e. $A_{\Delta}(0,\omega)E_{B}$
and the phase of the $T$-matrix, i.e. $\text{Arg}[-T]/\pi$ for different
average particle number {[}(a1), (a2), $\eta=0$; (b1),(b2), $\eta=0.5${]}
and different dissipation {[}$\gamma_{0}=0$: dotted red line; $\gamma_{0}=E_{F}$:
dashed blue line; $\text{\ensuremath{\gamma_{0}=30E_{F}}}$: black
sloid line{]} as a function of energy. The arrow indicate the bound
state $\omega_{b}^{inf}=0E_{F}$ and the threshold of the continue
state $\omega_{th}=2E_{F}$ for infinite loss rate case.In all the
subplots, we have set $m_{c}=m_{f}$, $E_{B}=2E_{F}$ and $k_{r}=k_{F}$
in the calculation.}
\end{figure}
\begin{figure*}[t]
\includegraphics[width=1\linewidth]{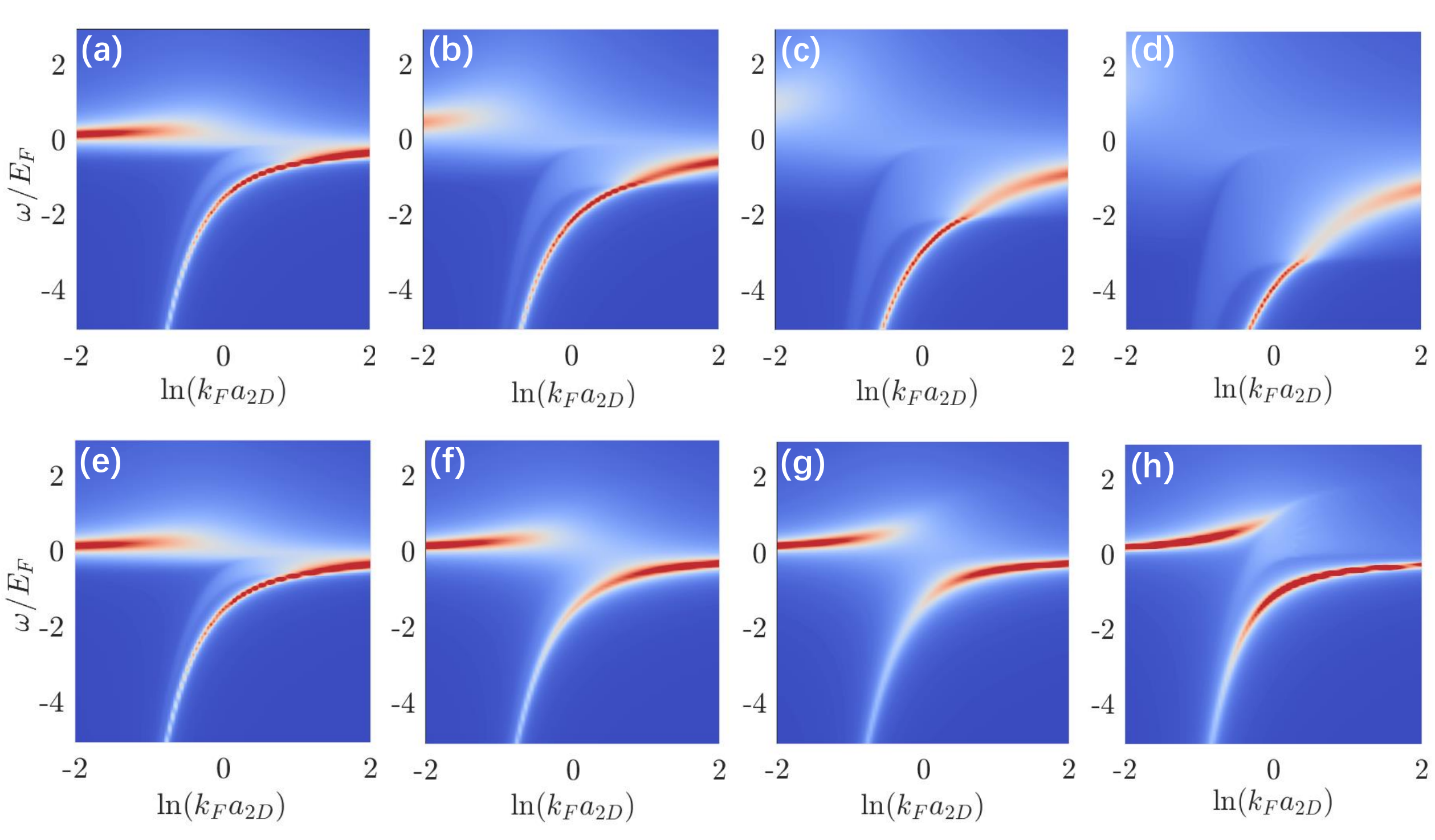}\caption{\label{Fig: polaron-spectral-fuction-interaction}(Color online) Polaron spectral function, i.e. $A_{c}(p=0,\omega)$ as a function of interaction
strength parameter, i.e. $\ln(k_{F}a_{2D})$ with different $k_{r}$
{[}(a), $k_{r}/k_{F}=1.0$; (b), $k_{r}/k_{F}=1.5$; (c), $k_{r}/k_{F}=2.0$;
(d), $k_{r}/k_{F}=2.5${]} for $\gamma_{0}=0.1E_{F}$, and with different
$\gamma_{0}$ {[}(e), $\gamma_{0}/E_{F}=0.1$; (f), $\gamma_{0}/E_{F}=1.0$;
(g), $\gamma_{0}/E_{F}=5.0$; (h), $\gamma_{0}/E_{F}=30${]} for $k_{r}/k_{F}=1.0$.
The average particle number is $\eta=0.5$.}
\end{figure*}

We further characterize the scattering by analyzing the phase of the
$T$-matrix. We start from the scenario in the absence of dissipation.
In the case of vaccum scattering where $\eta=0$, the many-body dynamics
reduces to a two-body one. The aforementioned dispersion of the pole
of $T_{0}$, i.e. $\omega_{0}$, generates an excitation with energy
$-E_{B}$ at zero momentum, which is confirmed by a sharp peak and
a $\pi$ jump, in the process of crossing the energy of bound state,
in Fig. \ref{Fig: molecule-phase}(a1) and (a2) respectively with
red dotted lines. The phase is zero in the regime without state, i.e.
$\omega<-E_{B}$, and jump to $\pi$ when crossing the poles of bound
states, i.e. $\omega=-E_{B}$. It maintains this value within the
gap between bound and continuous regime and then begins to decay monotonically
when $\omega>0$ in that the continuous state emerges. If $\eta$
is more than zero, the bath states within $k_{r}$ have a certain
probability of being occupied, then on average there is a finite density
of bath gases. Therefore, the formation of dressed molecular state
is more inclined to deduct the bath states within $k_{r}$, resulting
in a buleshift of bound state energy, as shown in Fig. \ref{Fig: molecule-dispersion}(e)
with blue dashed line whilst consistent with Fig. \ref{Fig: molecule-phase}(b1).
It is noted that although the continue state still arises at $\omega=0$,
which is the minimum of total kinetic energy of a molecule without
any constraints, the phase shows non monotonicity and has a peak at
$\omega>0$ (see details below).

When dissipation is applied to the background gas, the spectral function
of bound state exhibits greater broadening and blueshift than its
corresponding lossless case. Meanwhile, the $T$-matrix in the gap
regime is no longer a real one due to the finite dissipation, resulting
in the $\pi$ platform evolves into a broad peak (shown as blue dashed
line in Fig. \ref{Fig: molecule-phase}(a2) and (b2)). Interestingly,
whether the dissipation or the average particle number is not zero,
the phase of $T$-matrix is non monotonic in the continuous regime,
especially their peaks are in the same position which coincides with
that where the continuous state begins to emerge when the dissipation
is infinite. This commonality is due to the fact that the final effect
is to reconstruct the same “Fermi surface” no matter increasing the
dissipation strength or increasing the average number of particles.
In their respective limit cases, the effective ``Pauli blocking''
will push the threshold of continuous state to the optimal value obtained
in the subspace: $k_{f}>k_{r}$.

The numerical solutions with very large dissipation are also shown
by black solid lines in Fig. \ref{Fig: molecule-phase} for comparison
with the analytical results with infinite dissipation. It's clear
from the Eq. \eqref{eq:energy-infity} that the bound state at zero
momentum is $\omega_{b}^{inf}=\frac{\hbar^{2}k_{r}^{2}}{2m_{r}}-E_{B}$,
no matter $\eta$ is zero or finite, which is just at the position
of phase jump of the corresponding numerical results. To address the
threshold of continue state, we make use of the condition that the
bath atom can not be scattered into the dissipation subspace: $k_{f}<k_{r}$.
The momentum of bath atom is written as $\mathbf{k}_{f}=\frac{m_{f}}{M}\mathbf{q}-\mathbf{k}_{rel}$,
where $\mathbf{q}$ is the center of mass momentum and $\mathbf{k}_{rel}$
is the relative momentum. Then, the effective ``Pauli blocking''
requires
\begin{equation}
|\sqrt{2m_{r}E_{rel}}-\frac{m_{f}}{M}q|>k_{r},
\end{equation}
that is $E_{rel}>\omega_{th}-\frac{\hbar^{2}q^{2}}{2M}=\frac{(k_{r}+\frac{m_{f}}{M}q)^{2}}{2m_{r}}$.
Here the $\omega_{th}$ at $\mathbf{q}=0$ shows no difference with
the outcomes of numerical simulation with large dissipation, e.g.,
$\gamma_{0}=30E_{F}$.

\subsection{Polaron state}

\begin{figure}
\includegraphics[width=1\linewidth]{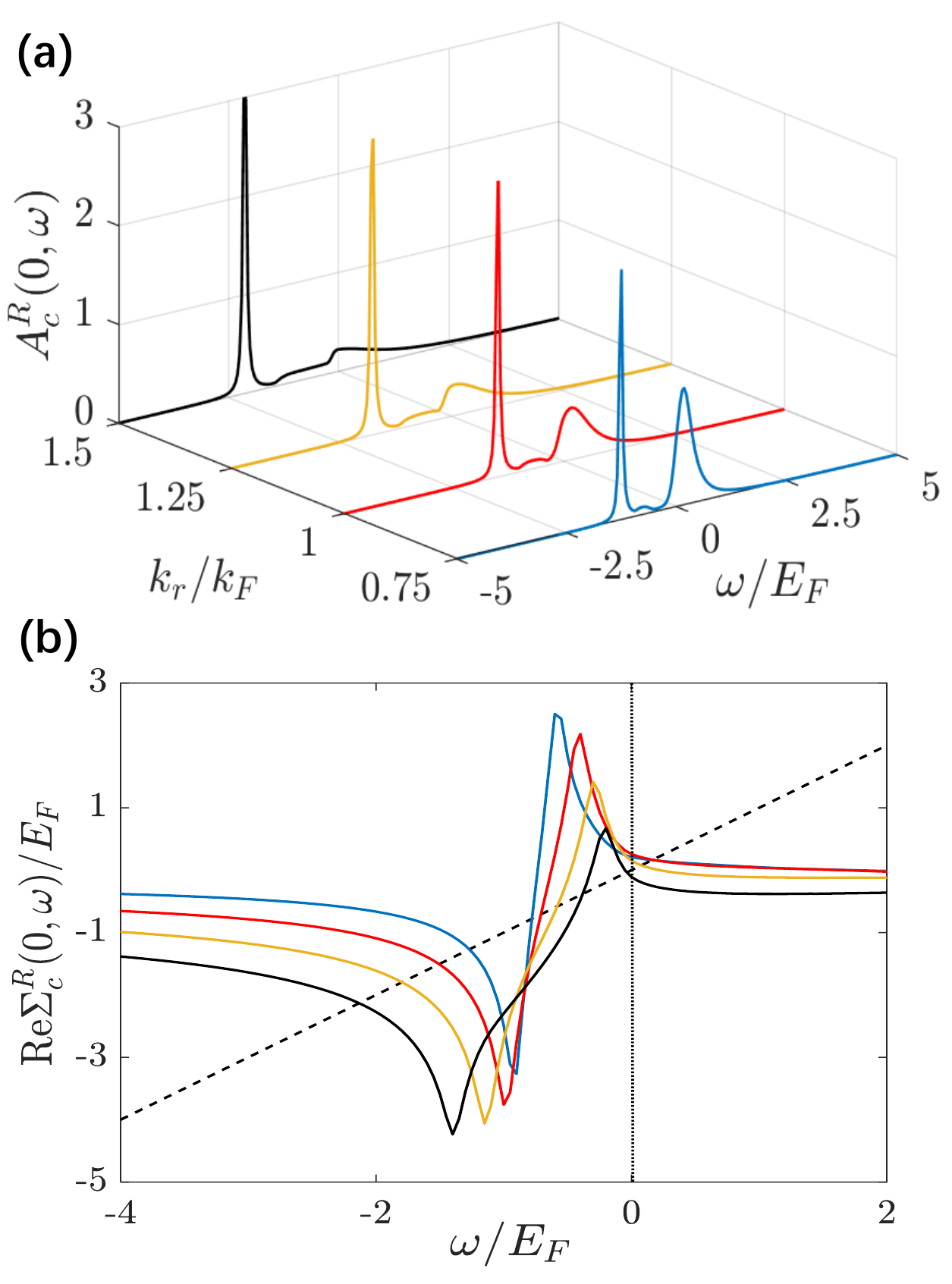}\caption{\label{Fig: polaron-spectral-function-kr}(Color online) Polaron spectral
function (a) and self-energy (b) with different $k_{r}$ {[}$k_{r}/k_{F}=0.75$:
blue line; $k_{r}/k_{F}=1.0$: red line; $k_{r}/k_{F}=1.25$: yellow
line; $k_{r}/k_{F}=1.5$: black line;{]}. Other parameters are: $E_{B}=2E_{F}$,
$\gamma_{0}=0.1E_{F}$, and $\eta=0.5$. The dashed black line is
the function of $g(\omega)=\omega$.}
\end{figure}

\begin{figure}
\includegraphics[width=1\linewidth]{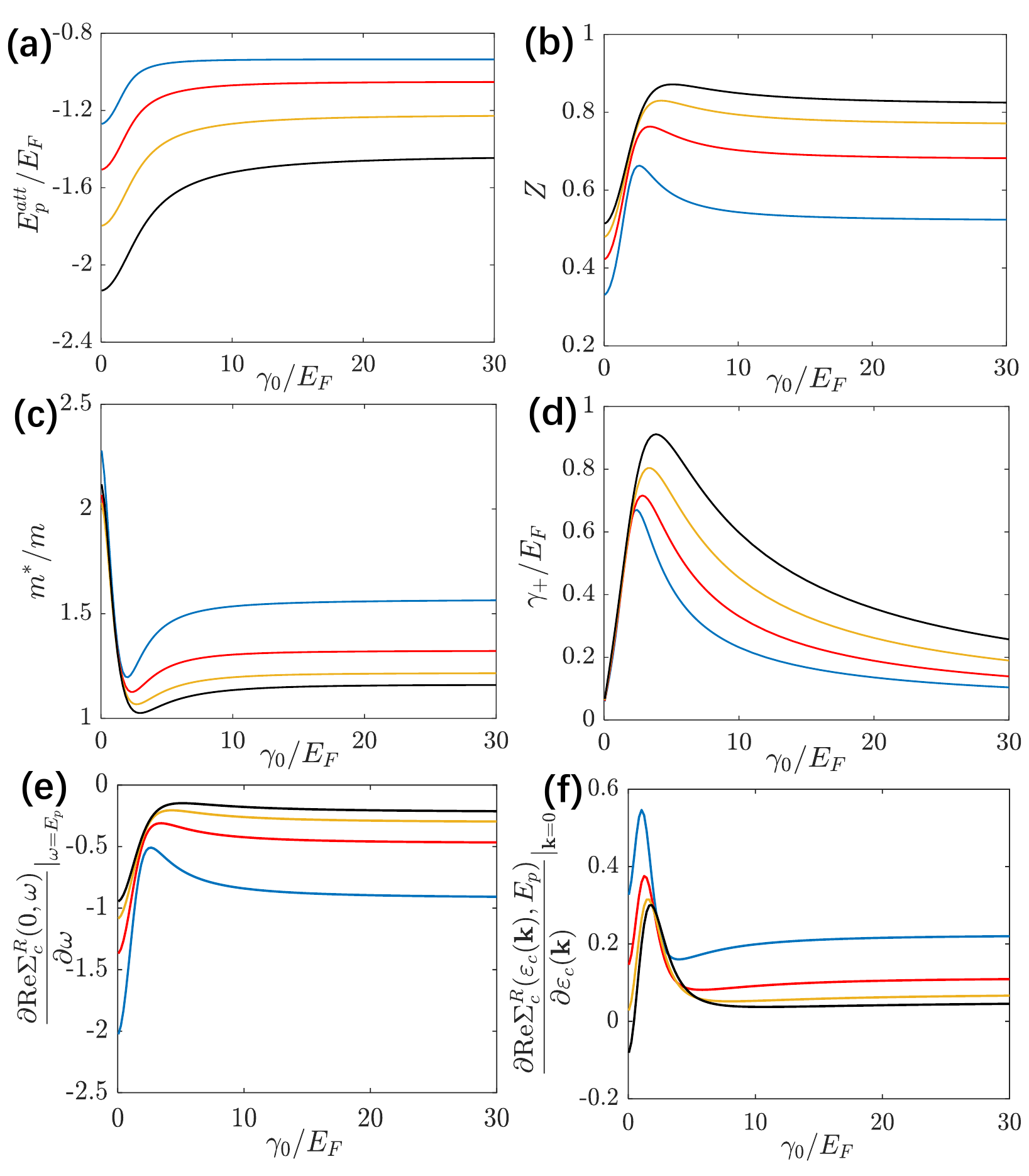}\caption{\label{Fig: four-quantities}(Color online) Quasiparticle parameters
and partial derivatives vs $\gamma_{0}$. Each color represents the
same $k_{r}$ as in Fig. \ref{Fig: polaron-spectral-function-kr},
so do $\eta$ and $E_{b}$.}
\end{figure}

In our system, the impurity does not directly participate in the nonunitary
evolution, but it is affected by dissipation through the scattering
with the intermediate molecular state. In this section, we mainly
concern the minimal energy state of polarons that occur at $\mathbf{k}=0$.
We can intuitively observe the excitations due to the spectral function
of impurity, the resonant peaks of which correspond to polaron states.
In Fig. \eqref{Fig: polaron-spectral-fuction-interaction}, we connect
$E_{B}$ with $a_{2D}$ by $E_{B}=1/ma_{2D}^{2}$ \cite{Demler2},
and present the contour plot of $A_{c}(\mathbf{k=}0,\omega)$. For
small $\gamma_{0}$ and $k_{r}=k_{F}$, the attractive branch is well-defined
with weak interaction. When the interaction increases, a repulsive
polaron emerges and the spectrum weight transfers from the attractive
polaron to repulsive polaron. Finallly the repulsive polaron becomes
long-lived and stable with large interaction. This scenario is basically
the same as when the background gases has Fermi surface \cite{Demler2}.
The polaron spectral function shows that they first diffuse and converge
with the increase of dissipation (see second row in Fig. \eqref{Fig: polaron-spectral-fuction-interaction}),
which can be attributed to the Zeno effect, as mentioned in the previous
section. When the $k_{r}$ exceeds $k_{F}$, the dissipation will
affect the higher energy excitations, which can be corroborated from
the obvious weakening of the repulsive polaron in Fig. \ref{Fig: polaron-spectral-fuction-interaction}(b),
(c), and (d) (see explanations below). In addition, the molecule-hole
continuum spread out more with the increase of $k_{r}$.

We can see the different responses of the two excitations to $k_{r}$
more clearly from the frequency sweep with a fixed $E_{b}$ in Fig.
\eqref{Fig: polaron-spectral-function-kr}(a). The attractive and
the repulsive polarons show different behaviors as $k_{r}$ on the
raise: although the energies of both of them are redshifted, the spectral
signal of the former is enhanced due to the weakening of the latter.
On the one hand, these frequency shifts in common can be confirmed
by calculating the positions of polaron states, which are obtained
by solving the Eq. \eqref{eq:polaron-energy} for $\mathbf{k}=0$
in our treatment (see the intersections of the straight dashed line
$g(\omega)=\omega$ and self-energy curves in Fig. \eqref{Fig: polaron-spectral-function-kr}(b)).
On the other hand, $k_{r}$ acts as Fermi surface to some extent.
In this picture, the particle-hole excitations around it dominate
the formation of polarons. Therefore, with the increase of $k_{r}$,
the repulsive polaron is easier to couple with the virtual molecular
state, that is, it will be more dissipated. Meanwhile, the attractive
branch becomes more and more like a bare impurity, which is reflected
in the increase of two-body decay and the decrease of effective mass,
as shown in Fig. \eqref{Fig: four-quantities}(c) and (d). As a result,
the spectral weight of the repulsive polaron is continuously shifted
to that of the attractive one.

To further describe the properties of attractive polaron and compare
them with the case without dissipation (at this time, the lifetime
of attractive polaron is infinite, i.e. $\gamma_{+}=0$), we present
polaron energy, quasi-particle residue, effective mass and two-body
decay rate as functions of $\gamma_{0}$ for mass balance case in
Fig. \eqref{Fig: four-quantities}. These quantities all initially
change vs $\gamma_{0}$ and gradually reach their saturation values
respectively, which should be explained by the Zeno effect as in the
previous section. The polaron energy displays redshift for specific
$\gamma_{0}$ with added $k_{r}$ showing consistence with Fig. \eqref{Fig: polaron-spectral-function-kr}.
The quasiparticle ratio $Z$, effective mass $m_{c}^{*}/m_{c}$ and
two-body decay $\gamma^{+}$ show nonmonotonic behavior vs $\gamma_{0}$.
The trend change positions of the three curves are basically the same,
and they are all at the dissipation of moderate intensity, i.e. $\gamma_{0}/E_{F}\sim2$
for our setup. In order to understand these results more clearly,
we exhibit two partial derivatives $\partial\text{Re}\Sigma_{c}^{R}(0,\omega)/\partial\omega|_{\omega=E_{p}}$
and $\partial\text{\text{Re}}\Sigma_{c}^{R}(\varepsilon_{c}(\mathbf{k}),E_{p})/\partial\varepsilon_{c}(\mathbf{k})|_{k=0}$
in Fig. \eqref{Fig: four-quantities}(e) and (f). Then, based on Eq.
\eqref{eq:Z} and \eqref{eq:m}, the behaviors of $Z$ and $m^{*}/m$
are unambiguous. These results are consistent with our point of view
in this paper, that is, only moderate dissipation is detrimental to
the formation of polarons. Since only the bath will be dissipated,
when $\gamma_{0}\sim E_{b}$ , the properties of the system are as
follows: 1) The proportion of impurity excitation in polaron goes
up; 2) The effective mass approaches unit; 3) The lifetime of polaron
is greatly compressed. These results are helpful for us to observe
attractive polaron in a dirty environment, e.g., a background gas
strongly coupled with itself environment.

\subsection{Experimental relevance}

\begin{figure}
\includegraphics[width=1\linewidth]{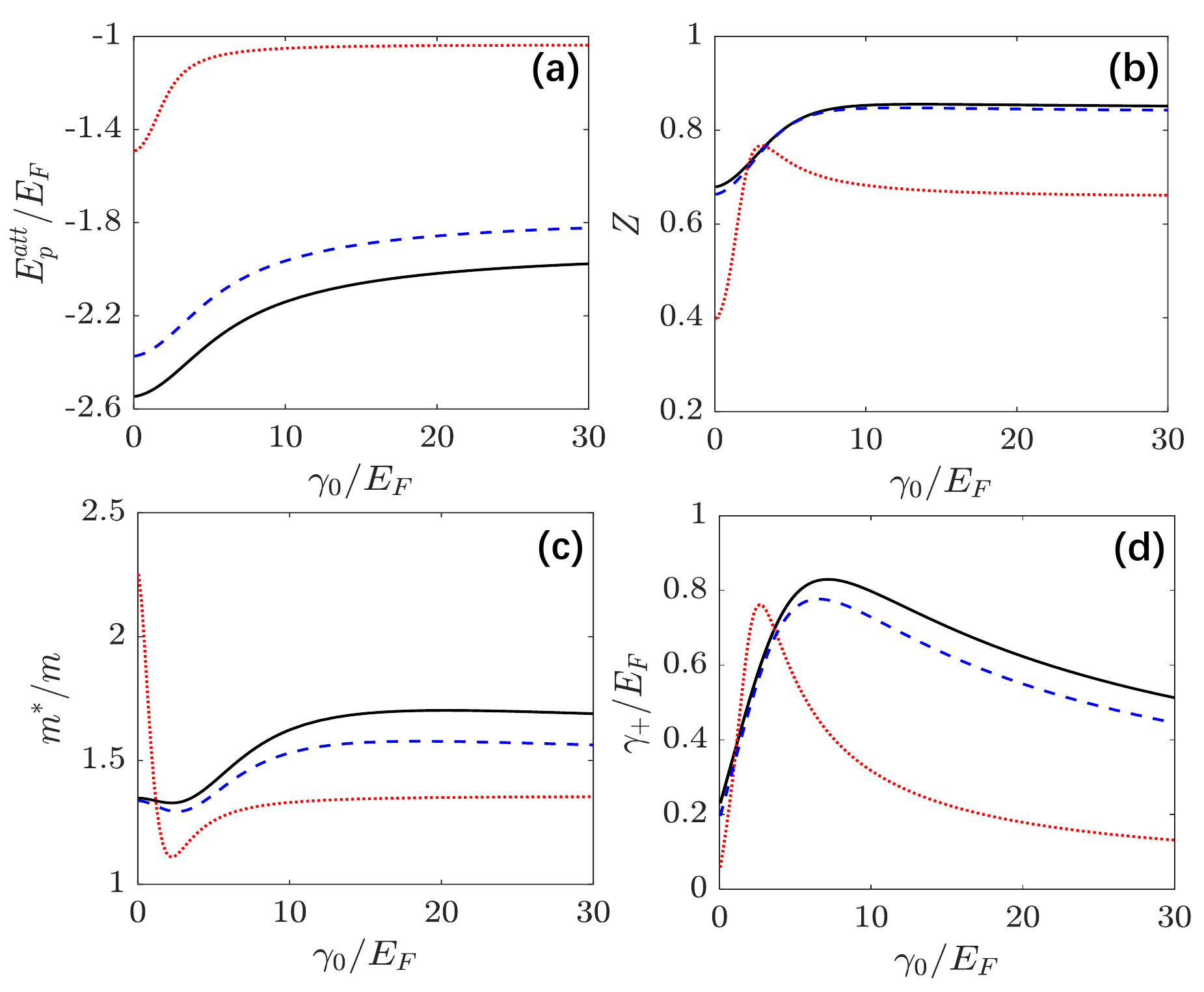}\caption{\label{Fig: experiment}(Color online) Polaron energy, residue, effective
mass, and two-body decay as a function of the dimensionless loss rate
$\gamma_{0}/E_{F}$ for the mixture of $^{6}$Li-$^{40}$K (black
solid line), $^{7}$Li-$^{40}$K (blue dashed line) and $^{7}$Li-$^{6}$Li
(red dashed line). The other parameters are set as $E_{B}=2E_{F}$,
$k_{r}=k_{F}$ and $\eta=0.5$.}
\end{figure}

Finally, we demonstrate the quasiparticle properties of the experiment
realizable systems. Since the impurity is a single atom, either a
Bose or Fermi impurity can be a candidate. Here, we consider one kind
of Fermi-Fermi and two kinds of Bose-Fermi mixtures as shown in Fig.
\eqref{Fig: experiment}. In the small impurity density limit, the
quasiparticle properties of the minority can be detected by RF spectrum
or Raman spectrum.

Comparing with the three kinds of experimental candidates, there is
a distinct change for the case with closer mass of impurity and bath
atom in terms of quasiparticle properties when we tune the loss rate.
So a mixture of minority $^{7}$Li and majority $^{6}$Li turn to
be a more desirable candidate for us to manipulate polaron state by
controlling the dissipation strength.

\section{\label{sec:Conclusion-and-outlook}Conclusion and outlook}

In summary, we use non-self-consistent $T$-matrix theory to solve
the polaron problem in a driven-dissipative bath. Non-equilibrium
Green's function is adopted to include the effect of the quantum jump
term in the LMEQ. We illustrate the non-trivial interplay among dissipation
range, dissipation strength, as well as interaction. In particular,
there are two analytic results: 1) we elucidate the mechanism that
the gap between molecular state and molecule-hole continuum decreases
monotonically with the dissipation range when the dissipation strength
is very small. 2) we obtain the dispersion of bound state as a function
of interaction strength and dissipation range when the dissipation
strength is infinit large, which is independent of the average number
of particles in the background medium. We numerically demonstrate
the molecular and polaron spectrum functions with different dissipation
strength, and find that their resonant peaks become diffuse only under
moderate dissipation, indicating the quasiparticles cease to exist
in this regime. Meanwhile, the positions of both the attractive and
repulsive polarons display redshift when the dissipation range increases,
and the two polarons exhibit opposite behavior in this process: we
see an increment and a decrement of the weights of attrative and repulsive
polarons respectively. We also calculate the quasiparticle parameters
vs dissipation strength with different dissipation ranges for the
attractive polaron. The residue, effective mass and two-body decay
show nonmonotonic characters due to the interplay between the intrinsic
energy scale of the system and the measurement frequency from the
environment (understanding dissipation from another perspective).
However, they eventually tend to their respective saturation values
in the quantum-Zeno limit, i.e. $\gamma_{0}=\infty$. Our model can
be implemented in recent experiments \cite{Luo,Gadway,Gerbier,Takahashi},
and the results in this work can help us to find appropriate setups
in dissipative cold atom experiments to observe clear-cut polaron
signals.

Based on the technical methods used in this work, we can further study
the experimental realizable system, e.g., the exciton-polariton system
in quantum wells embedded in an optical micro cavity. The leakage
of photons from the cavity and the decay of exciton via radiative
and non-radiative processes make the platform a natural open quantum
system \cite{Deng,Dang,West}. How to use nonequilibrium Green's
function to characterize natural open systems is reserved for future
research.

\section{Acknowledgment}

Y.C. was supported by the NSFC-China (Grants No. 11704029 and No.
12174024). J.Z. was supported by the NSFC-China (Grants No. 11504038).
The authors acknowledge stimulating discussions with Hui Hu and Jia
Wang.

\end{document}